\documentclass[twocolumn,showpacs,preprintnumbers,amssymb]{revtex4}
\usepackage{graphicx, epsfig, epstopdf} 
\usepackage{amsmath, amsfonts, amssymb}
\usepackage{bm, array}
\usepackage[usenames]{color}
\usepackage[
      colorlinks=true,
      linkcolor=blue,
      urlcolor=blue,
      filecolor=blue,
      citecolor=red,
      pdfstartview=FitV,
      pdftitle={},
      pdfauthor={},
      pdfsubject={},
      pdfkeywords={},
      pdfpagemode=None,
      bookmarksopen=true
]{hyperref}

\def\be{\begin{equation}}
\def\ee{\end{equation}}
\def\beq{\begin{eqnarray}}
\def\eeq{\end{eqnarray}}

\begin{document}

\centerline{}
\title{Collapsing rotating shells in Myers-Perry-AdS$_5$ spacetime:
A perturbative approach}

\author{
Jorge V. Rocha,$^{1}$
\footnote{Electronic address: jorge.v.rocha@tecnico.ulisboa.pt}
Raphael Santarelli,$^{2}$
\footnote{Electronic address: santarelli@ifsc.usp.br}
T\'erence Delsate,$^{1,3}$
\footnote{Electronic address: terence.delsate@umons.ac.be}
}
\affiliation{${^1}$ CENTRA, Deptartamento de F\'{\i}sica, Instituto Superior T\'ecnico, Universidade de Lisboa, Av.~Rovisco Pais 1, 1049 Lisboa, Portugal}
\affiliation{${^2}$ Instituto de F\'isica de S\~ao Carlos, Universidade de S\~ao Paulo, Caixa Postal 369, CEP 13560-970, S\~ao Carlos, SP, Brazil}
\affiliation{${^3}$ Theoretical and Mathematical Physics Department,  University of
Mons, UMONS, 20, Place du Parc, B-7000 Mons, Belgium}

\date{\today}

\begin{abstract}
We study gravitational perturbations sourced by a rotating test shell collapsing into five-dimensional Myers-Perry black holes in anti-de Sitter (AdS). Our attention is restricted to the case in which the two possible angular momenta of the geometry are set equal. In this situation the background is cohomogeneity-1, which amounts to a crucial technical simplification.
It is found that the linearized Einstein equations are consistent only when the test shell is corotating with the spacetime. However, it is argued that this is a consequence of the matter on the shell being described by dust or, more precisely, noninteracting test particles. We compute the mass and angular momenta of the perturbed spacetime using a counterterm subtraction method, for which we provide an explicit formula that has not appeared previously in the literature. The results are in agreement with the expected expressions for energy and angular momenta of geodesic particles in AdS$_5$.
\end{abstract}


\maketitle



\section{Introduction}
\label{sec:Intro}

Addressing rotation in the context of general relativity is a notoriously difficult problem. Even when considering an isolated object, rotation deforms bodies away from sphericity, explicitly introducing dependence on polar angles. Conjugated with the nonlinearity of the theory, this basic fact hampers attempts to analytically solve the field equations unless some solution-generation technique is applicable. Just to give an example from black hole (BH) physics, it took roughly half a century to discover the rotating generalization~\cite{Kerr:1963} of the four-dimensional static black hole. Even when considering the linearization around a given rotating solution, the issue of its stability is far from trivial~\cite{Whiting:1988vc}.

Nevertheless, rotation obviously plays a crucial role in astrophysical systems.
In the axisymmetric gravitational collapse of stars the presence of an angular momentum can generate centrifugal forces strong enough to prevent the formation of a black hole, leading instead to a bounce. This phenomenon has been demonstrated numerically~\cite{Nakamura:1981,Stark:1985da}; generically there is little analytic control over such collapses. A derivation of this effect, by focusing on the dynamics on the equatorial plane, was reported several years before in Ref.~\cite{Wagoner:1965} but it made use of some not entirely justified assumptions.

There is by now a vast literature on the subject of gravitational collapse but it is almost entirely dedicated to spherically symmetric scenarios (see, e.g.~\cite{Joshi:2012mk}). The situation concerning collapse with rotation is much less developed. A very useful approach, due to its simplicity, is to consider collapsing {\em shells} of matter but, even so, the inclusion of rotation typically impedes a full analytic treatment of the problem. A few exceptions are provided by Refs.~\cite{Cohen:1968,Lindblom:1974bq}, which rely on a slow rotation approximation. More recently, an interesting study by Mann {\it et al.}~\cite{Mann:2008rx} was able to tackle the problem {\em exactly}, at the expense of considering three spacetime dimensions.

In this paper we investigate gravitational perturbations induced by a collapsing shell of test particles into a rotating black hole in five dimensions, with a negative cosmological constant. This is the same approach adopted in~\cite{Rocha:2011wp}, where perturbations of the three-dimensional rotating black hole in anti-de Sitter (AdS) by an in-falling circular ring of test particles were examined. Here, however, we shall perturb the BH spacetime with a continuous test {\em shell} preserving all the angular symmetry of the five-dimensional background.
In higher dimensions there can be several independent angular momenta~\cite{footnote1}. We take advantage of the fact that, in odd dimensions ($D\geq5$), when all the angular momenta are equal --- and we restrict to this case --- the rotating black hole geometry is cohomogeneity-1~\cite{Kunduri:2006qa}, allowing us to describe the solution by functions of only one coordinate. The black hole event horizon and all constant radial surfaces are topologically odd-dimensional spheres. Such spaces can be endowed with metrics that break isotropy but preserve homogeneity and this was fruitfully used in Ref.~\cite{Bizon:2005cp} to study vacuum gravitational collapse with rotation but in a numerically less-demanding scenario. 
For simplicity, we will consider the five-dimensional case, but, as discussed in Sec.~\ref{sec:conc}, we expect that our results may generalize to higher odd dimensions without much effort. 

Our study can be regarded as an extension of an analysis by Zerilli (see Appendix G of Ref.~\cite{Zerilli:1971wd}) to higher dimensions and by the addition of two more parameters (rotation and the cosmological constant). In the nonrotating case the metric perturbation can be expanded in higher $D$ spherical harmonics, which all decouple in the linearized Einstein equations, and our results describe the solution of the lowest harmonic, i.e., monopole perturbation. However, for nonvanishing spin of the background, the equations no longer decouple and our solution cannot be expected to describe the projection of the perturbed metric onto the lowest angular harmonic (zero mode) when the source breaks the angular symmetry of the background.

The motivation for such a study is at least twofold. First, the analysis of linearized perturbations allows us, in principle, to identify the ``energy'' and ``angular momentum'' of a rotating test shell in five-dimensional, asymptotically AdS spacetimes. This in turn determines the variation of the charges of the spacetime caused by each point particle. Such knowledge is essential to perform tests of the cosmic censorship conjecture~\cite{Penrose:1969pc} in such spacetimes, by attempting to overspin extremal rotating black holes with point particles as envisaged originally by Wald~\cite{Wald:1974} and extended recently to higher dimensions in Ref.~\cite{BouhmadiLopez:2010vc}, and to spacetimes with a cosmological constant in Refs.~\cite{Rocha:2011wp, Zhang:2013tba, Rocha:2014}.

On the other hand, gravitational collapse in AdS spacetimes has important implications for thermalization of strongly coupled conformal field theories (CFTs) via the gauge-gravity duality (see~\cite{Aharony:1999ti} for a review). In Ref.~\cite{Bhattacharyya:2009uu} the fast thermalization process of a CFT (on a sphere) perturbed away from the vacuum by a homogeneous short pulse was inferred by following the spherically symmetric collapse of a scalar field into a black hole in the holographic dual setting (see~\cite{Garfinkle:2011hm,Bantilan:2012vu} for related numerical works). It is naturally desirable to know whether the inclusion of rotation has some impact on this picture, as all of these studies rely on spherical symmetry.
We take a first step in this direction by computing the one-point function of the holographically dual stress-energy tensor, following the counterterm subtraction proposal of Ref.~\cite{Balasubramanian:1999re}.

\bigskip

The outline of the paper is as follows. In the next section we briefly present the black hole geometries we shall consider. In Sec.~\ref{sec:perturb} we discuss the linearized Einstein equations, the geodesics on this spacetime, and the stress-energy tensor the test particles generate. These linearized perturbation equations are then solved in Sec.~\ref{sec:solving} and the result is used to determine the variation of the mass and angular momentum of the spacetime caused by the test particles. We conclude in Sec.~\ref{sec:conc} with some discussion and remarks. The appendixes gather several technical details concerning gauge fixing and solving the gravitational perturbation equations.

\section{Equally spinning black holes in five dimensions
\label{sec:BHgeometry}}

In this section we gather the relevant details about five-dimensional rotating black holes with a negative cosmological constant when both angular momenta are equal. These spacetimes will serve as a background on top of which we analyze linear perturbations in Sec.~\ref{sec:solving}.
These black hole solutions in AdS$_5$ were first presented in~\cite{Hawking:1998kw} and have been extended to arbitrary dimensions $D\geq4$ in~\cite{Gibbons:2004js,Gibbons:2004uw}, thus generalizing the well-known Myers-Perry family~\cite{Myers:1986un} to include a cosmological constant.

In five spacetime dimensions one can pick two orthogonal planes of rotation. In general the solution is parametrized by two independent angular momenta $a_1$ and $a_2$, in addition to a mass parameter $M$. As shown in~\cite{Kunduri:2006qa}, when the rotation parameters are set equal, $a_1=a_2=a$, the isometry group gets enhanced and the solution becomes cohomogeneity-1. The geometry essentially depends on a single radial coordinate only, and the metric can be written as follows:
\beq
 ds^2  &=&  - f(r)^2 dt^2 + g(r)^2 dr^2 + r^2 \widehat{g}_{ab} dx^a dx^b \nonumber \\
  && +\, h(r)^2 \left[ d\psi + A_a dx^a - \Omega(r) dt \right]^2\,,
\label{eq:metric}
\eeq
where
\beq
g(r)^2  &=&  \left( 1 + \frac{r^2}{\ell^2} - \frac{2GM\Xi}{r^2} + \frac{2GMa^2}{r^4} \right)^{-1}\,, \label{eq:metricfuncs1}\\
h(r)^2  &=&  r^2 \left( 1 + \frac{2GMa^2}{r^4} \right)\,, \qquad \Omega(r) =  \frac{2GMa}{r^2 h(r)^2}\,, \label{eq:metricfuncs2}\\
f(r)  &=&  \frac{r}{g(r) h(r)}\,, \qquad \Xi = 1 - \frac{a^2}{\ell^2}\,. \label{eq:metricfuncs3}
\eeq
In the above expressions $\widehat{g}_{ab}$ represents the Fubini-Study metric on the complex projective space $CP^1$, which is isomorphic to the sphere $S^2$, and $A=A_a dx^a$ is its K\"ahler potential:
\be
\widehat{g}_{ab} dx^a dx^b  =  \frac{1}{4} \left( d\theta^2 + \sin^2\theta \, d\phi^2  \right)\,, \;
A = \frac{1}{2} \cos\theta \, d\phi\,.
\ee

The above form of the metric~\eqref{eq:metric} extends to all higher odd dimensions $D=2N+3$~\cite{Kunduri:2006qa}: this is made possible by the fact that the sphere $S^{2N+1}$ can be written as an $S^1$ bundle over $CP^N$. For the case $D=5$ on which we concentrate, this corresponds to the familiar Hopf fibration. The coordinate $\psi$ parametrizes the $S^1$ fiber and has period $2\pi$. The two orthogonal rotation planes correspond to $\theta=0$ and $\theta=\pi$ in these coordinates, i.e., the rotation planes are mapped to the poles of the $S^2$.

This metric is a solution of the Einstein equations with a negative cosmological constant,
\be
R_{\mu\nu} = - 4\ell^{-2} g_{\mu\nu}\,.
\ee
The largest real root, $r_+$, of $g^{-2}$ marks an event horizon that possesses the geometry of a homogeneously squashed $S^3$ (written above as its Hopf fibration). The mass ${\cal M}$ and angular momentum ${\cal J}$ of the spacetime are given by~\cite{Kunduri:2006qa}
\be
{\cal M} = \frac{\pi M}{4G} \left( 3 + \frac{a^2}{\ell^2} \right)\,, \qquad
{\cal J} = \frac{\pi M a}{G} \,.
\ee
%

%
%

In the expressions above and in the rest of the manuscript, we will use natural units, normalizing the speed of light to $c=1$, but we shall explicitly keep factors of the Newton constant $G$.

\section{Linearized gravitational perturbations
\label{sec:perturb}}

We are interested in investigating the consequences of perturbing the background spacetime by a (comparatively light) in-falling membrane of test particles homogeneously distributed on the squashed $S^3$. The meaning of this will be made precise in Sec.~\ref{sec:conserv}. For now we just remark that this situation preserves the full rotational symmetry of the background~\cite{footnote2}, and, in particular, the equal angular momentum property. Of course, stationarity is lost due to the presence of the in-falling test particles.

For this study we will adopt the framework of linearized perturbations.
The perturbed metric is obtained from the background metric $g_{\mu\nu}$ by
\be
\widetilde{g}_{\mu\nu} = g_{\mu\nu} + h_{\mu\nu} \,.
\ee
As mentioned previously, the background metrics we shall consider are solutions
of the sourceless cosmological Einstein equations in five dimensions and reads
\be
G^c_{\mu\nu} \equiv G_{\mu\nu}-\frac{6}{\ell^2} g_{\mu\nu}=0\,.
\label{eq:EinstEqs}
\ee
Here $G_{\mu\nu}=R_{\mu\nu}-\frac{1}{2}g_{\mu\nu}R$ denotes the Einstein tensor, $\Lambda \equiv -4/\ell^2$ represents the (negative) cosmological constant and $\ell$ is the AdS radius.
The linearized perturbation equations derived from~\eqref{eq:EinstEqs} are the following~\cite{Ortin:2004ms, Murata:2008xr}:
\begin{align}
2\delta G^c_{\mu\nu} \equiv & -\nabla^2h_{\mu\nu} +2\nabla^\lambda \nabla_{(\mu}h_{\nu)\lambda} -\nabla_\mu \nabla_\nu h  \nonumber\\
	& +g_{\mu\nu} \left(\nabla^2 h - \nabla^\alpha \nabla^\beta  h_{\alpha\beta} \right) -h_{\mu\nu}R  \nonumber\\
  & +g_{\mu\nu} h_{\alpha\beta}R^{\alpha\beta} - 12\ell^{-2} h_{\mu\nu} \nonumber\\
  =& 16\pi G\, T_{\mu\nu}\,,
\label{eq:linEinstein}
\end{align}
where covariant derivatives are taken with respect to the background metric and $h\equiv g^{\mu\nu}h_{\mu\nu}$ denotes the trace of the metric perturbation.
$T_{\mu\nu}$ is the stress-energy tensor of the test particles that will drive the perturbation.
By virtue of the Bianchi identity, the stress-energy tensor must be divergenceless to ensure consistency of equations~\eqref{eq:linEinstein}.
This occurs if and only if the source particles follow geodesics. These
equations easily generalize to higher spacetime dimensions.

Particularizing to the background vacuum solution~\eqref{eq:metric}, the
linearized perturbation equations become
\begin{align}
-\nabla^2h_{\mu\nu} &+2\nabla^\lambda \nabla_{(\mu}h_{\nu)\lambda} -\nabla_\mu \nabla_\nu h  \nonumber\\
	+& g_{\mu\nu} \left(\nabla^2 h - \nabla^\alpha \nabla^\beta  h_{\alpha\beta} \right) +8\ell^{-2}\left(h_{\mu\nu} -\frac{1}{2}g_{\mu\nu} h \right)  \nonumber\\
    &= 16\pi G\, T_{\mu\nu}\,.
\label{eq:linEinstein2}
\end{align}
In Sec.~\ref{sec:Tmunu} we will determine the generic form of the stress-energy tensor for a shell of test particles that preserves all of the angular isometries of the background.

As usual, diffeomorphism invariance of the theory implies the existence of gauge freedom that can be used to eliminate some terms appearing in the differential operator on the left-hand side of Eq.~\eqref{eq:linEinstein2}. Typically, the choice of transverse traceless gauge is made, reducing the differential operator to the Lichnerowicz operator.
%
%
However, the expected form of the perturbation, i.e. one leading to a shift of the mass and rotation parameters
\be
M\rightarrow M+\delta M\,, \qquad  a\rightarrow a+\delta a\,,
\ee
in the region outside the shell is not consistent with such gauge fixing. We will find it more convenient to make a different gauge choice, one that preserves the symmetry of the background. This is discussed in Appendix~\ref{sec:gauge}.

\subsection{Conserved quantities and geodesics
\label{sec:conserv}}

Consider a test particle moving along some geodesic of the spacetime~\eqref{eq:metric}. The world line is described by $z^\mu(\tau)=(T(\tau),R(\tau),\Psi(\tau),\Theta(\tau),\Phi(\tau))$, where $\tau$ is an affine parameter.
A rotating stationary spacetime specified by a metric tensor $g_{\mu\nu}$ possesses both time-like and rotational Killing vectors, from which we can build three conserved quantities,
\begin{flalign}
E &\equiv -g_{\mu\nu} {\partial_t}^\mu \dot z^\nu = -\frac{dT}{d\tau} \! \left[ g_{tt}+g_{t\psi}\frac{d\Psi}{dT}+g_{t\phi}\frac{d\Phi}{dT} \right], \label{eq:EandL1}\\
L_\psi &\equiv g_{\mu\nu} {\partial_\psi}^\mu \dot z^\nu = \frac{dT}{d\tau}\left[ g_{t\psi}+g_{\psi\psi}\frac{d\Psi}{dT} +g_{\psi\phi}\frac{d\Phi}{dT} \right], \label{eq:EandL2}\\
L_\phi &\equiv g_{\mu\nu} {\partial_\phi}^\mu \dot z^\nu = \frac{dT}{d\tau}\left[ g_{t\phi}+g_{\psi\phi}\frac{d\Psi}{dT} +g_{\phi\phi}\frac{d\Phi}{dT} \right], \label{eq:EandL3}
\end{flalign}
where the dot indicates derivation with respect to $\tau$.

Now, note that 
\beq
g_{t\phi} &=& \frac{\cos\theta}{2} g_{t\psi}\,, \qquad  g_{\psi\phi} = \frac{\cos\theta}{2} g_{\psi\psi}\,, \nonumber\\
g_{\phi\phi} &=& \frac{\cos\theta}{2} g_{\psi\phi} + \frac{r^2 \sin^2\theta}{4}\,,
\eeq
so a test particle whose motion lies entirely in the rotation plane $\theta=0$ has $L_\phi = \frac{1}{2} L_\psi$.
Similarly, a test particle whose motion lies entirely in the rotation plane $\theta=\pi$ has $L_\phi = -\frac{1}{2} L_\psi$.
These geodesics satisfy
\be
\dot{\Theta}(\tau) = \dot{\Phi}(\tau) = 0\,.
\label{eq:geodesics}
\ee

\begin{figure*}
\centering
\includegraphics[width=\textwidth]{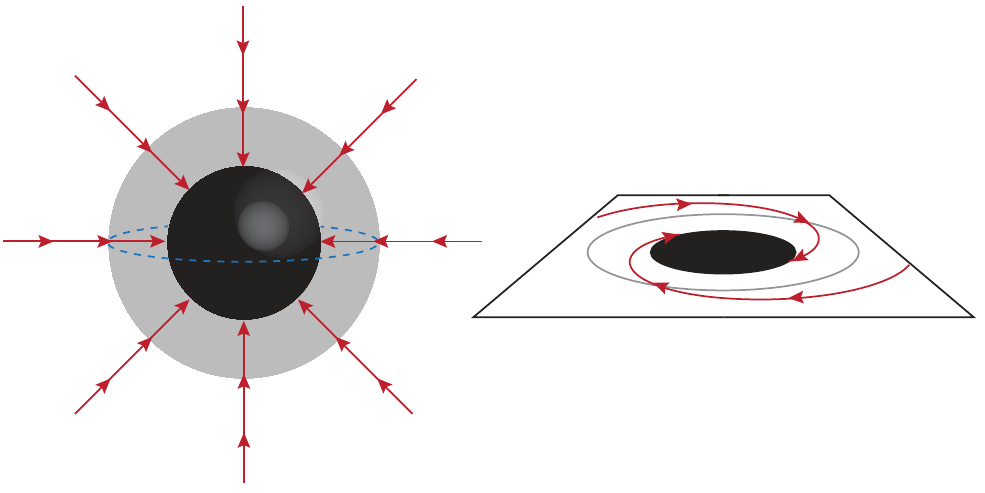}
\caption{Geodesics followed by individual test particles shown in arrowed red lines. The left panel shows the point particle trajectories projected onto the three-dimensional space parametrized by spherical coordinates $(r,\theta,\phi)$. The trajectories in this space are radial. The right panel illustrates similar trajectories now projected onto the two-dimensional space parametrized by polar coordinates $(r,\psi)$. In both panels the location of the black hole event horizon, in black, and of the continuous shell of test particles at a snapshot in time, in grey, are superimposed.}
\label{fig:geodesics}
\end{figure*}

In fact, there exist geodesics obeying~\eqref{eq:geodesics} {\em for any value of $\theta$ and $\phi$}.
Such geodesics simply correspond to static trajectories on the $S^2$ and they have
\be
L_\phi= \frac{\cos\theta}{2}L_\psi \,.
\ee
Thus, to preserve the most amount of symmetry of the background, in this work we will consider a membrane of test particles homogeneously smeared over the $S^2$, with each point particle infalling radially.
The particles will also be homogeneously distributed over the $S^1$ but they will possess some rotation along the $\psi$ direction (see Fig.~\ref{fig:geodesics}).

The remaining equations governing the test particle trajectories are easily obtained by inverting~\eqref{eq:EandL1} and~\eqref{eq:EandL2} and using $g_{\mu\nu}\dot z^\mu \dot z^\nu=-\epsilon$ to obtain the radial equation
\beq
\dot{T}     &=& \frac{E- \Omega L_\psi}{f^2}\,,\label{eq:timeequation}\\
\dot{\Psi}  &=& \frac{h^2\Omega E + \left( f^2-h^2\Omega^2 \right)L_\psi}{h^2f^2}\,,\\
\dot{R}^2   &=& - \frac{ \epsilon - f^{-2}\left[ E-\Omega L_\psi \right]^2 +h^{-2}L_\psi^2}{g^2}\,,\label{eq:radialequation} 
\eeq
where $\epsilon=1,0$ for time-like or null geodesics, respectively.
In these equations the metric functions $f,g,h,\Omega$ should all be considered as functions of $R(\tau)$ instead of $r$.

In this work we will restrict to the case $\epsilon=0$, corresponding to null geodesics that reach the time-like boundary of AdS. Although it is not immediate how to separately define the black hole and the test particle when the latter cannot be moved all the way out to time-like infinity, one can nevertheless compute the effect that the introduction of a test particle in the background geometry has on the energy and angular momenta of the spacetime. This will be done in Sec.~\ref{sec:solving}.

\subsection{The stress-energy tensor
\label{sec:Tmunu}}

For a single point particle of rest mass $m_0$, the stress-energy tensor is given by~\cite{Zerilli:1971wd}
\be
T^{\mu\nu}_{(pp)} = m_0 \int \delta^{(5)}(x-z(\tau))\frac{dz^{\mu}}{d\tau}\frac{dz^{\nu}}{d\tau}d\tau\,,
\label{eq:Tmunu}
\ee
where, recall, $\tau$ denotes an affine parameter along the worldline $z^{\mu}(\tau)$.

As mentioned before, we will perturb the background spacetime with test particles homogeneously (and continuously) distributed over both the $S^1$ and $S^2$. The stress-energy tensor will thus be a function of the coordinates $t$ and $r$ only, and it can be computed by a simple change of variables. After performing the time integral and then integrating over the angular coordinates (to yield $T^{\mu\nu}_{(pp)}$ smeared over the squashed $S^3$) we obtain~\cite{Zerilli:1971wd, Lousto:1996sx}
\be
T^{\mu\nu} = 4m_0\vartheta^{\mu\nu} \frac{\delta(r-R(t))}{r^3}\,,
\quad  \vartheta^{\mu\nu} \equiv \frac{dT}{d\tau}\frac{dz^{\mu}}{dt}\frac{dz^{\nu}}{dt}\,.
\label{eq:shellTmunu}
\ee
For massive particles the quantity $m_0$ appearing in~\eqref{eq:shellTmunu} is the mass density of the test shell. For massless particles the form~\eqref{eq:Tmunu} for the stress-energy tensor is still valid~\cite{Ortin:2004ms} but the parameter $m_0$ should then be interpreted as an energy density throughout the shell.
The covariant components of the stress-energy tensor may be expressed as follows:
\begin{flalign}
\vartheta_{tt} &= - E \left[ g_{tt}+g_{t\psi}\frac{d\Psi}{dt} \right], \quad \;\;
\vartheta_{tr} = - E\, g_{rr} \frac{dR}{dt}\,, \nonumber\\
\vartheta_{t\psi} &= - E \left[ g_{t\psi}+g_{\psi\psi}\frac{d\Psi}{dt} \right], \quad
\vartheta_{rr} =  \frac{dT}{d\tau} g_{rr}^2 \left(\frac{dR}{dt}\right)^2, \nonumber\\
\vartheta_{\psi\psi} &=  L_\psi \left[ g_{t\psi}+g_{\psi\psi}\frac{d\Psi}{dt} \right], \quad \,
\vartheta_{r\psi} =  L_\psi g_{rr} \frac{dR}{dt}\,, \label{eq:stresstensor} \\
\vartheta_{t\phi} &= - E \left[ g_{t\phi}+g_{\psi\phi}\frac{d\Psi}{dt} \right], \quad \,
\vartheta_{r\phi} =  L_\phi g_{rr} \frac{dR}{dt}\,, \nonumber\\
\vartheta_{\psi\phi} &=  L_\psi \left[ g_{t\phi}+g_{\psi\phi}\frac{d\Psi}{dt} \right], \quad \;\,
\vartheta_{\phi\phi} =  L_\phi \left[ g_{t\phi}+g_{\psi\phi}\frac{d\Psi}{dt} \right]. \nonumber
\end{flalign}
The components $\vartheta_{\mu\theta}\,$ all vanish. This stress-energy tensor is traceless (${T_\mu}^\mu=0$) and conserved ($\nabla_\mu T^{\mu\nu}=0$), as is imposed by the Bianchi identities. Note that using the relations
\beq
-E \left[ g_{t\psi}+g_{\psi\psi}\frac{d\Psi}{dt} \right] &=& L_\psi \left[ g_{tt}+g_{t\psi}\frac{d\Psi}{dt} \right]\,, \\
-E \left[ g_{t\phi}+g_{\psi\phi}\frac{d\Psi}{dt} \right] &=& L_\phi \left[ g_{tt}+g_{t\psi}\frac{d\Psi}{dt} \right]\,, \\
L_\psi \left[ g_{t\phi}+g_{\psi\phi}\frac{d\Psi}{dt} \right] &=& L_\phi \left[ g_{t\psi}+g_{\psi\psi}\frac{d\Psi}{dt} \right]\,,
\eeq
one can express $\vartheta_{t\psi}$ in terms of $L_\psi$ and both $\vartheta_{t\phi}$ and $\vartheta_{\psi\phi}$ in terms of $L_\phi$.

\section{Solving the linearized perturbation equations
\label{sec:solving}}

In this section we will solve the perturbation equations~\eqref{eq:linEinstein2}, with the stress-energy tensor given by Eq.~\eqref{eq:shellTmunu}.

To this end, we begin by gauge fixing so that the metric perturbation $h_{\mu\nu}$ takes a convenient form, similar to that dictated by a (time-independent) shift of mass and angular momentum. Several perturbation components automatically vanish, which simplifies the process of solving the perturbation equations. Having a solution, we may then infer the effect the test shell has on the spacetime, namely by determining the change in the mass $\delta {\cal M}$ and in the angular momentum $\delta {\cal J}$ imputed on the background. This was essentially the same strategy followed in the much simpler $(2+1)$-dimensional case~\cite{Rocha:2011wp}. A similar approach is followed in Ref.~\cite{Murata:2008xr}, though with a different problem in mind.

We will first consider the fully nonrotating case, where the background is static and the test shell infalls radially. This is a instructive limit because it can be solved fully analytically and builds up some intuition for the rotating case, which we address in a separate subsection.

In Appendix~\ref{sec:gauge} it is shown that the metric perturbation can be gauge fixed into the following form, possessing only four independent components:
\be
h_{\mu \nu}= \left[
 \begin{array}{ccccc}
  h_{tt} \;      & 0                       & h_{t\psi} \;    & 0 \; & \frac{\cos\theta}{2} h_{t\psi} \\
  0                & g(r)^4 h_{rr} \; & 0                   & 0    & 0   \\
  *                 & 0                       & h_{\psi\psi} & 0    & \frac{\cos\theta}{2} h_{\psi\psi} \\
  0                & 0                       & 0                   & 0    & 0 \\
  *                 & 0                       & *                   & 0    & \frac{\cos^2\theta}{4} h_{\psi\psi} \\  
\end{array}\right] \,.
\label{eq:gaugefix}
\ee
A factor of $g(r)^4$ is inserted in the $\{rr\}$ component for convenience.

Therefore, we have a total of fifteen equations to solve for only four functions (of $t$ and $r$). Not all of Eqs.~\eqref{eq:linEinstein2} are linearly independent. In fact, there is only one dynamical equation that determines one of the nontrivial perturbation components, say $h_{\psi\psi}$. The other components are fixed by constraint equations. This is in agreement with the appendix of reference~\cite{Murata:2008xr}.

\subsection{The nonrotating case
\label{sec:Non-rotating}}

In the nonrotating case ($a=0$ and $L_\psi=0$) one can in fact gauge fix so that $h_{t\psi}=h_{\psi\psi}=0$ (see Appendix~\ref{sec:gauge}). The metric perturbation has only two nontrivial components: $h_{tt}$ and $h_{rr}$.

As mentioned above, in this special case one can solve the linearized perturbation equations exactly. The system of fifteen equations reduces to just four equations, corresponding to components $\{tt\}, \{tr\}, \{rr\}$,
\begin{widetext}
\begin{flalign}
&\frac{3 \left(r^4+\ell^2 r^2 -2 M \ell^2 \right)}{\ell^2 r^5}
\partial_r \left[ r^2 h_{rr} \right]
= \frac{64\pi\,m_0 E \left(r^4+\ell^2 r^2 -2 M \ell^2 \right)}{\ell^2 r^5}\delta(r-R(t))\,, \label{eq:sys_eqs_1}\\
&\frac{3 \ell^2 r}{r^4+\ell^2 r^2 -2 M \ell^2} \partial_t h_{rr}
 = -\frac{64\pi\,m_0 E\, \ell^2 R'(t)}{r \left(r^4+\ell^2 r^2 -2 M \ell^2 \right)}\delta(r-R(t))\,, \label{eq:sys_eqs_2}\\
& 3\ell^2
\left\{ \frac{2 r^2 \left(\ell^2+2 r^2\right)}{\left(r^4+\ell^2 r^2 -2 M \ell^2 \right)^2} h_{rr} + \frac{1}{r} \partial_r \left[ \frac{r^2 h_{tt}}{r^4+\ell^2 r^2 -2 M \ell^2 } \right]  \right\} = - \frac{64\pi\,m_0 \ell^4 r R'(t)^2 \dot{T}}{\left(r^4+\ell^2 r^2 -2 M \ell^2 \right)^2}\delta(r-R(t))\,, \label{eq:sys_eqs_3}
\end{flalign}
in addition to a longer and not so enlightening equation from the $\{\psi\psi\}$ component, which we avoid presenting. The remaining equations are either trivially satisfied or linearly dependent on the former. It is straightforward to solve the system~(\ref{eq:sys_eqs_1}-\ref{eq:sys_eqs_3}) and the final result is
\beq
h_{tt}(t,r) &=& \frac{C_1}{r^2} - \frac{r^4+\ell^2r^2-2 M \ell^2}{3 \ell^2 r^2}C_2(t) + \frac{64 \pi\,m_0 E}{3 r^2} \Theta(r-R(t)) \left[1 - 2 \frac{ r^4+\ell^2r^2-2M\ell^2 }{ R(t)^4+\ell^2R(t)^2-2M\ell^2 } \right], \label{eq:sol_htt1}\\
h_{rr}(t,r) &=& \frac{C_1}{r^2} + \frac{64 \pi\,m_0 E}{3 r^2} \Theta(r-R(t))\,. \label{eq:sol_hrr1}
\eeq
\end{widetext}
Here, $C_1$ is a constant and $C_2$ represents an arbitrary function of $t$. We have used Eq.~\eqref{eq:timeequation} to replace $\dot{T}$ in~\eqref{eq:sys_eqs_3}. It can be checked that this solution also satisfies the remaining independent equation arising from the $\{\psi\psi\}$ component. The solution for the asymptotically flat case can be easily obtained by taking the $\ell\to\infty$ limit.

Some comments regarding the solution~(\ref{eq:sol_htt1}-\ref{eq:sol_hrr1}) are now in order. The terms not proportional to $m_0$ are obviously not sourced by the test particles. The terms proportional to $C_1$ correspond to a simple (static) perturbation of the mass parameter. On the other hand, the time-dependent term $C_2(t)$ can be gauged away by using the residual gauge freedom discussed in Appendix~\ref{sec:gauge}. Thus we may set $C_1=C_2(t)=0$.
The component $h_{rr}$ then takes exactly the form that would be expected: it is proportional to $r^{-2}$ and to the Heaviside function $\Theta(r-R(t))$.
The $\{tt\}$ component, on the other hand, is more involved, though still proportional to $\Theta(r-R(t))$. It appears to grow as $r^2$, but this is again a gauge artifact. The residual gauge freedom can also be used to eliminate the entire second term inside the square brackets in Eq.~\eqref{eq:sol_htt1}, at the expense of introducing some nontrivial behavior for $r<R(t)$. Stated differently, if we completely gauge fix by requiring the metric perturbations to decay as $r\to\infty$, then $h_{tt}=h_{rr}=O(r^{-2})$ for $r>R(t)$. In that case, $h_{tt}=h_{rr}$ for $r>R(t)$ although $h_{tt}$ then acquires nontrivial support inside the test shell.

Nevertheless, it is the $h_{rr}$ component that determines the mass of the spacetime. This can be computed using the quasilocal stress tensor formalism of Ref.~\cite{Balasubramanian:1999re}, which relies on a counterterm subtraction method. For an asymptotically AdS$_5$ spacetime the required counterterms to render the boundary stress tensor finite have been identified in~\cite{Balasubramanian:1999re}, to which we refer for details of the computation. A long but straightforward calculation reveals that the variation in the mass of the spacetime is given by
\begin{flalign}
\delta{\cal M} &= \lim_{r\to\infty} \frac{3}{64\pi G} \int_0^{2\pi} d\phi \int_0^{2\pi} d\psi \int_0^{\pi} d\theta \, \sin\theta \nonumber\\
& \quad \times \left[r^2 h_{rr} - (\ell^2+8M) \frac{\ell^2 h_{tt}}{8r^2} \right]  
= \frac{8\pi^2 m_0 E}{G} \,,
\label{deltaM1}
\end{flalign}
i.e. it is determined by the net energy $m_0 E$ of the test particles (the factor of $8\pi^2$ accounts for the integration of the continuous distribution over the 3-sphere).

We conclude that the test shell in this nonrotating setting has the effect (at linear level) of adding an energy $8\pi^2m_0E/G$ to the background spacetime.

\bigskip

\subsection{The rotating case
\label{sec:Rotating}}

In the fully rotating case we have four functions of $t$ and $r$ to solve for. The linearized Einstein equations are very lengthy but can be reduced to a system of four equations (see Appendix~\ref{sec:equations}): a decoupled second order partial differential equation (PDE) for $h_{\psi\psi}$, a coupled ordinary differential equation (ODE) that determines $h_{t\psi}$ once $h_{\psi\psi}$ is given, and two constraint equations that fix $h_{rr}$ and $h_{tt}$ once the other two nontrivial components are known.
In the rotating case the non-homogeneous PDE for $h_{\psi\psi}$ does not seem possible to solve in closed form. Nevertheless, one can find a solution in the form of an asymptotic expansion in powers of $1/r$ and $1/R(t)$ near the boundary of AdS. This is in fact all we need to compute the conserved charges of the test shell.

Given the general form of the differential equations being solved --- with source terms proportional to delta distributions and step functions --- we seek solutions of the form
\be
h_{\mu\nu}(t,r)=\eta_{\mu\nu}(R(t),r) \Theta(r-R(t))\,,
\label{eq:ansatz}
\ee
for some tensor field $\eta_{\mu\nu}(R(t),r)$.

This is also what is expected on physical grounds: perturbations preserving the
full $U(2)$ spatial symmetry of the background vanish inside the shell, modulo a
residual gauge freedom similar to the nonrotating case. Inserting this ansatz
in Eqs.~(\ref{hpsipsi}-\ref{htt}) we obtain a system of differential equations
coming from the terms proportional to $\Theta(r-R(t))$, which are supplemented
by a set of boundary conditions stemming from the terms proportional to
$\delta(r-R(t))$ and $\delta'(r-R(t))$ (see Appendix~\ref{sec:checks}).

Let us focus on the decoupled equation~\eqref{hpsipsi} governing the component $h_{\psi\psi}$. Application of the ansatz~\eqref{eq:ansatz} returns a long nonhomogeneous second order PDE for $\eta_{\psi\psi}$. It can be shown that 
\be
\eta_{\psi\psi}^{(p)}(r) = \frac{32\pi m_0 \ell^2 a \left(2aE-3 L_{\psi }\right)}{\left(a^2-3 \ell^2\right) r^2}
\ee
is a particular solution of this nonhomogeneous equation. A close examination reveals that the associated homogeneous equation is separable, $\eta_{\psi\psi}(R,r)=\Sigma(R)\sigma(r)$. The solutions for each of the two functions $\sigma$ and $\Sigma$ can be expressed in an asymptotic power series in $1/r$ and $1/R(t)$, respectively, and the result is
\begin{widetext}
\begin{flalign}
\eta_{\psi\psi}^{(h)}(R,r) = & \int _0^{\infty } d\kappa\; \alpha(\kappa)
\Big[ \frac{1}{r^2} +\frac{\kappa ^2 \ell^4}{12r^4} +\frac{-128 a^2 M+384 \ell^2 M-28 \ell^6 \kappa ^2+\ell^8 \kappa ^4}{384r^6} + O(r^{-8}) \Big] \nonumber\\
& \times \Big[1+\frac{\kappa^2 E^2 \ell^6}{2(E^2 \ell^2-L_{\psi}^2)R^2} +\frac{\kappa^2 E^2 \ell^8 \left(E^2 \ell^2 (\ell^2 \kappa^2-8)+12 L_{\psi}^2\right)}{24(E^2 \ell^2-L_{\psi}^2)^2R^4} + O(R^{-6}) \Big]\,.
\end{flalign}
Here, the integration variable $\kappa$ is the separability constant and the integration stems from the basic fact that any linear combination of solutions --- parametrized by $\kappa$ --- will also give a solution to the homogeneous equation. Defining
\be
A_i \equiv \int _0^\infty d\kappa\; \kappa^{2i}\alpha(\kappa)\,, \qquad i=0,1,2,\dots
\ee
we may write the general asymptotic solution as
\begin{flalign}
\eta_{\psi\psi}(R,r) =
& \frac{32\pi m_0 \ell^2 a \left(2aE-3 L_{\psi }\right)}{\left(a^2-3 \ell^2\right) r^2} + \frac{A_0}{r^2} + \frac{A_1 \ell^4}{12r^4} + \frac{A_1 E^2 \ell^6}{2(E^2 \ell^2-L_{\psi}^2)r^2 R^2} - \frac{A_0(a^2-3 \ell^2) M}{3r^6} - \frac{7A_1 \ell^6}{96r^6} \nonumber\\
& + \frac{A_2 \ell^8}{384r^6} + \frac{A_2 E^2 \ell^{10}}{24(E^2 \ell^2-L_{\psi}^2)r^4 R^2} - \frac{A_1 E^2 \ell^8 \left(2 E^2 \ell^2-3 L_{\psi}^2\right)}{6 (E^2 \ell^2-L_{\psi}^2)^2 r^2 R^4} + \frac{A_2 E^4 \ell^{12}}{24 (E^2 \ell^2-L_{\psi}^2)^2 r^2 R^4} + \dots\;,
\label{eq:asy_eta}
\end{flalign}
\end{widetext}
where the dots refer to higher order terms in $1/r$ and $1/R$ that when evaluated at $r=R(t)$ become $O(R^{-8})$. 

The constraint~\eqref{BC1} yields the simple boundary condition $L_\psi^2 \eta_{\psi\psi}(R,R)=0$. If $L_\psi\neq0$ this fixes the parameters $A_i$ to be
\begin{flalign}
A_0 &= -\frac{32\pi a\ell^2  m_0 \left(2aE-3 L_{\psi }\right)}{\left(a^2-3 \ell^2\right)}\,, \\
A_1 &= 0\,, \\
A_2 &= -\frac{4096\pi a M  m_0 \left(2aE - 3L_{\psi}\right) (E^2 \ell^2-L_{\psi}^2)^2}{\ell^6 \left(33 E^4 \ell^4-18 E^2 \ell^2 L_{\psi}^2+L_{\psi}^4\right)}\,.
\end{flalign}
Thus, the solution at this point takes the simpler form
\beq
\eta_{\psi\psi}(R,r) &=& -\frac{512\pi a M\ell^4 m_0 E^2 }{3 r^6 R^4}  \left(r^2-R^2\right) \left(2 a E-3 L_{\psi }\right) \nonumber\\
&& \times
\frac{E^2 \ell^2 \left(r^2+2 R^2\right)-R^2 L_{\psi }^2}{33 E^4 \ell^4-18 E^2
\ell^2 L_{\psi}^2+L_{\psi}^4} + \ldots\,.
\label{eq:sol_etapsipsi}
\eeq
To complete the determination of $h_{\psi\psi}$ we must also impose the boundary conditions~\eqref{BC2} and~\eqref{BC3}. Plugging the solution~\eqref{eq:sol_etapsipsi} in either~\eqref{BC2} or~\eqref{BC3} returns an equation of the form
\be
m_0 L_\psi^2 \, Y(R) = 0\,, 
\ee
where $Y(R)$ is an unenlightening rational function of $R$. So the boundary conditions can only be satisfied --- for generic values of the radial position of the test shell --- if $L_\psi=0$ (assuming $m_0\neq0$, otherwise there would be no test particles).
In this case the asymptotic solution~\eqref{eq:asy_eta} satisfying boundary conditions~\eqref{BC2} and~\eqref{BC3} turns out to be
\be
\eta_{\psi\psi}(R,r) \simeq - \frac{1024 \pi a^2 M \ell^2 m_0 E \left(r^4+r^2R^2-2R^4\right)}{99r^6R^4}\,,
\ee
which can equally be obtained by taking the limit $L_\psi\to0$ in~\eqref{eq:sol_etapsipsi}.

We conclude that, in general, the (hyperbolic) PDE for the component $h_{\psi\psi}$ has no nontrivial solution satisfying the boundary conditions. The only consistent case is $L_\psi=0$, for which an asymptotic solution can indeed be
found. In other words, \textit{a shell of perfect fluid dust preserving the full
spatial symmetry of the background but not co-rotating with the spacetime is
inconsistent with the linearized Einstein equations}. Thus, we restrict to
$L_\psi=0$ in the remainder of this section. It is easy to see from
Eq.~\eqref{eq:radialequation} that in this case the motion of the (null) test
shell indeed describes a full collapse, i.e., a bounce never occurs.

Following the same strategy, we can now straightforwardly compute the remaining nontrivial metric perturbation components by integrating Eqs.~(\ref{htpsi}-\ref{htt}). In doing so, we must keep terms up to the order $O(r^{-6})$, including terms up to order $r^{-4}O(R(t)^{-2})$, $r^{-2}O(R(t)^{-4})$, $O(R(t)^{-6})$, and $r^{2}O(R(t)^{-8})$, which contribute at the same order when imposing the boundary conditions at $r=R(t)$. The final result is
\begin{widetext}
\begin{flalign}
h_{\psi\psi}(t,r) \simeq& - \frac{1024 \pi a^2 M \ell^2 m_0 E}{99} \Theta(r-R(t)) \frac{r^4+r^2R(t)^2-2R(t)^4}{r^6R(t)^4}\,, \label{eq:sol_hpsipsi2}\\
h_{t\psi}(t,r) \simeq& \frac{128 \pi a M \ell^2 m_0 E}{3} \Theta(r-R(t)) \frac{r^4-R(t)^4}{r^2R(t)^8}\,, \label{eq:sol_htpsi2}\\
h_{rr}(t,r) \simeq& \frac{64 \pi\,m_0 E}{3 r^2} \Theta(r-R(t)) \left[1 + 32\, a^2 M\frac{2r^2+3R(t)^2}{99\, r^2 R(t)^4} \right], \label{eq:sol_hrr2}\\
h_{tt}(t,r) \simeq& \frac{64 \pi\,m_0 E}{3 r^2} \Theta(r-R(t)) \left[1 - 2 \frac{r^4+\ell^2 r^2-2M\ell^2}{R(t)^4} 
+ \frac{a^2 M}{99} \left( \frac{16}{R(t)^4}-\frac{34}{r^4}-\frac{396r^2(r^2+\ell^2)}{R(t)^8} \right) \right]. \label{eq:sol_htt2}
\end{flalign}
\end{widetext}
This is in full agreement with the exact result obtained for the nonrotating case: this asymptotic solution reduces to Eqs.~\eqref{eq:sol_htt1} and~\eqref{eq:sol_hrr1} when $a=0$ (up to higher order corrections in powers of $R(t)^{-1}$). Observe that we cannot take the flat limit $\ell\to\infty$ consistently from the above asymptotic solution: terms of subleading order in powers of $R(t)$ --- which are being discarded in Eqs.~(\ref{eq:sol_hpsipsi2}--\ref{eq:sol_htt2}) --- will generically introduce extra factors of $\ell$.

In writing the above solution we have set some integration constants to zero in order to retain only the part sourced by the test shell that interests us. Expressions~\eqref{eq:sol_hpsipsi2} and~\eqref{eq:sol_hrr2} explicitly show that $h_{\psi\psi}$ and $h_{rr}$ both decay as $~r^{-2}$. As in the nonrotating case, here we can also resort to the residual gauge freedom to eliminate the $O(r^2)$ terms that appear in ~\eqref{eq:sol_htpsi2} and~\eqref{eq:sol_htt2}, at the expense of introducing nontrivial behavior for $r<R(t)$ [see Appendix~\ref{sec:gauge}]. It turns out that after performing this residual gauge fixing the component $h_{t\psi}$ becomes of the order $O(r^{-4})$.

Adopting the quasilocal stress tensor approach of~\cite{Balasubramanian:1999re}, these linear perturbations determine the variation of the mass and angular momentum of the spacetime, which are given by
\begin{flalign}
\delta{\cal M} &= \lim_{r\to\infty} \frac{1}{64\pi G} \int_0^{2\pi} d\phi \int_0^{2\pi} d\psi \int_0^{\pi} d\theta \, \sin\theta \nonumber\\
& \quad \times  \left[ 3r^2 h_{rr} + \frac{2r^2}{\ell^2} h_{\psi\psi} - \frac{r^3}{\ell^2}\partial_r h_{\psi\psi} \right. \nonumber\\
& \qquad \left. - \frac{3\ell^4+24\ell^2M+8a^2 M}{8r^2} h_{tt} \right]  
= \frac{8\pi^2 m_0 E}{G} \,,   \label{deltaM2}  \\
\delta{\cal J}_\psi &= \lim_{r\to\infty} \frac{1}{64\pi G} \int_0^{2\pi} d\phi \int_0^{2\pi} d\psi \int_0^{\pi} d\theta \, \sin\theta \nonumber\\
& \qquad \times \left[ 2r^2 h_{t\psi} - r^3 \partial_r h_{t\psi} \right]  = 0\,. \label{deltaJ}
\end{flalign}

Note that the result for the increment in mass is the same as in the nonrotating case, Eq.~\eqref{deltaM1}: the finite terms (in the limit $r\to\infty$) coming from $h_{rr}$ that are proportional to $a^2 M$ exactly cancel the contribution from $h_{\psi\psi}$. The variation in the angular momentum vanishes, as expected, since we considered a corotating test shell. To be precise, the quantity computed in Eq.~\eqref{deltaJ} is the change in the $\psi-$component of the angular momentum of the spacetime --- the only component that is initially nonzero. The variations of the other two angular components trivially vanish.

In the spirit of holographic renormalization, the bulk coordinates are naturally split as $x^\mu=(r,x^i)$, with $x^i=(t,\psi,\theta,\phi)$, and the radial coordinate $r$ is interpreted roughly as an inverse energy scale from the dual field theory point of view. The quantities computed above are simply proportional to the $\{tt\}$ and $\{t\psi\}$ components of the one-point function of the (perturbed) dual stress-energy tensor. More generally, the expectation value of the boundary stress-energy tensor is dictated, via the quasilocal stress tensor ${\cal T}_{ij}$, by the asymptotic beha\-vior of the metric~\cite{Balasubramanian:1999re, Bantilan:2012vu}:
\be
\left< T_{ij}(x^k) \right> = \lim_{r\to\infty} r^2 {\cal T}_{ij}(r,x^k)\,.
\ee
We refer the reader to~\cite{Balasubramanian:1999re} for the explicit expression of the quasilocal stress tensor.

For completeness, we present the results for the perturbation induced by the collapsing test shell on the one-point function of the boundary stress-energy tensor:
\begin{flalign}
& \left< \delta T_{tt} \right> = \frac{4m_0 E}{G\ell}\,,\\
& \left< \delta T_{\psi\psi} \right> = \frac{4m_0 E \ell}{3G} \left(1-\frac{128a^2 M}{99R(t)^4}\right)\,,\\
& \left< \delta T_{\psi\phi} \right> = \frac{\cos\theta}{2} \left< \delta T_{\psi\psi} \right>\,,\\
& \left< \delta T_{\theta\theta} \right> = \frac{m_0 E \ell}{3G} \left(1+\frac{64a^2 M}{99R(t)^4}\right)\,,\\
& \left< \delta T_{\phi\phi} \right> = \frac{\cos^2\theta}{4} \left< \delta T_{\psi\psi} \right>+\sin^2\theta \left< \delta T_{\theta\theta} \right> \,.
\end{flalign}
All other components of $\left<\delta T_{ij}\right>$ not related by symmetry vanish. This calculation also uncovers an interesting effect: although the actual radial location of the test shell does not influence the energy and momentum density, there is an explicit dependence on $R(t)$ in the spatial components, which corresponds to pressures and shear.

\section{Discussion
\label{sec:conc}}

In this work we have studied the collapse of a rotating shell of null test particles towards a five-dimensional rotating black hole in asymptotically AdS spacetime with equal rotation parameters. We employed a perturbative approach by considering the aforementioned black hole geometry as a fixed background, and studying the (linearized) effects of a test shell preserving all the rotational symmetry of the spacetime.

We first considered the nonrotating case, which is instructive because the linearized Einstein equations can be solved exactly. The presence of the shell increases the total energy of the spacetime by an amount precisely equal to the mass of the shell. 
For the fully rotating case, we were only able to solve the perturbation equations asymptotically. Nevertheless, this was sufficient to show that the introduction of a continuous and homogeneous distribution of (noninteracting) test particles preserving the $U(2)$ symmetry of the equally rotating Myers-Perry-AdS$_5$ background is only consistent if the angular momentum parameter of the particles vanishes,  $L_\psi=0$, i.e., if the shell is corotating with the background.

In practice, the noninteracting character of the test particles constituting
the shell, which is sourcing the perturbations, translates into the shell equation of state being that of dust.
The above conclusion implies that, if we want to
consider a general $L_\psi \neq 0$ case, we must add extra terms to the
stress-energy tensor 
to alter the angular momentum of the black hole.
Roughly speaking, the perfect fluid form of the stress-energy tensor does not carry the appropriate charge to affect the spin of the black hole, and consistency of the linearized Einstein equations for $L_\psi\neq0$ requires the introduction of additional forces. The angular momentum of the test shell should be given by an off diagonal element of the stress-energy tensor, namely $T^{t\psi}$. So one expects that at least some momentum flux is needed for consistency. This is indeed confirmed by an exact treatment that will be presented elsewhere~\cite{DRS}.

It seems very likely that our results for the five-dimensional equally spinning Myers-Perry-AdS black hole generalizes to all odd higher dimensions, since the structure of the linearized equations remains unaltered. In particular, there should still be four independent equations governing the perturbations, even though the total number of degrees of freedom grows as $D^2$. This is a consequence of the high degree of symmetry we impose on the perturbed spacetime.
The specific case we studied has equal angular momenta in the two independent rotation planes but in less symmetric situations we expect similar results~\cite{footnote3}.
In particular, a thin dust shell in such a spacetime must be corotating; otherwise, it must be composed of a nonperfect fluid.


As a byproduct of our studies, we derived explicit expressions for the change in mass and angular momentum of these spacetimes induced by the test shell though the gravitational perturbations it sources. The counterterm prescription we adopted~\cite{Balasubramanian:1999re}, also known as the quasilocal stress tensor formalism, reproduces the expected results. 
In any case, the expressions obtained, Eqs.~\eqref{deltaM2} and~\eqref{deltaJ}, are generically valid for gravitational perturbations of these spacetimes that preserve the background angular symmetry and its AdS asymptotics.
This provides a hint that this procedure may also give consistent results in a nonlinear situation, e.g., a backreacted collapse or BH collision in AdS, provided the gauge is suitably chosen, namely that the spacetime is asymptotically locally AdS.

A point worth mentioning is related to the explicit time dependence in the asymptotic solution found for the rotating case, Eqs.~(\ref{eq:sol_hpsipsi2}-\ref{eq:sol_htt2}). As discussed in~\cite{Bizon:2005cp,Bizon:2006wk}, even though the full angular symmetry of the background is preserved by the perturbations considered, the Birkhoff-Jebsen theorem can be evaded in odd dimensions. Therefore, the spacetime outside the test shell need not be static, or even stationary, as it turns out to be the case.

In this investigation we have resorted to gauge fixing in order to solve the perturbation equations. 
Alternatively, we could have chosen to work with a gauge-invariant formulation, as done in Ref.~\cite{Murata:2008xr} to study sourceless perturbations. It would be interesting to extend that technology to nonvacuum perturbations.

\section*{Acknowledgements}

We thank Vitor Cardoso and Jan Steinhoff for useful discussions.
We also thank the anonymous referee for a helpful suggestion.
J.~V.~R. is supported by {\it Funda\c{c}\~ao para a Ci\^encia e Tecnologia} (FCT)-Portugal through Contract No.~SFRH/BPD/47332/2008. R.~S. is supported by Grant No.~2012/20039-6, S\~ao Paulo Research Foundation (FAPESP). R.~S. also thanks \textit{Centro Multidisciplinar de Astrof\'isica} and \textit{Instituto Superior T\'ecnico} (IST) for hospitality during the initial phase of this work.

\appendix

\section{Gauge transformations
\label{sec:gauge}}

It is convenient to decompose the metric perturbation into scalars, vectors and tensors according to their transformation properties under a general transformation of the angular coordinates $\{\psi,\theta,\phi\}$.
Thus we have four scalar components $h_{AB}$, two vectors $h_{Ai}$ and one tensor $h_{ij}$, where $A,B=t,r$ and $i,j=\psi,\theta,\phi$.
Each of these sectors can be expanded in terms of Wigner functions $D^J_{KM}(\psi,\theta,\phi)$, see reference~\cite{Murata:2008xr}. In five dimensions there exist two commuting angular momentum operators whose Casimir operators coincide.
The Wigner functions are the eigenfunctions of these operators.

Given the high degree of symmetry of the matter distribution we are assuming, namely $U(1)\times SU(2)$, the induced metric perturbation will preserve the symmetry of the background.
This implies that only the zero-mode ($J=M=K=0$) of the expansion in Wigner functions can be excited and these modes decouple from the rest.

According to~\cite{Murata:2008xr} there are only seven independent components in the perturbation zero-mode,
\be
\{ h_{tt}\,,   h_{tr}\,,  h_{rr}\,,  h_{t\psi}\,,  h_{r\psi}\,,  h_{\psi\psi}\,,  h_{\theta\theta} \}\,,
\ee
while the four components $h_{\mu\theta}$ with $\mu\neq\theta$ vanish identically and the remaining components are given by
\beq
h_{t\phi}&=&\frac{\cos\theta}{2}h_{t\psi}\,,  \quad    h_{r\phi}=\frac{\cos\theta}{2}h_{r\psi}\,, \\
h_{\psi\phi}&=&\frac{\cos\theta}{2}h_{\psi\psi}\,, \quad  h_{\phi\phi}=\sin^2\theta\, h_{\theta\theta} + \frac{\cos^2\theta}{4}h_{\psi\psi}\,. \nonumber
\eeq

Under an infinitesimal coordinate transformation the metric perturbation transforms as
\be
h_{\mu\nu} \longrightarrow h_{\mu\nu}^{\rm new} = h_{\mu\nu} + \nabla_{(\mu}\xi_{\nu)}\,.
\ee
In order not to spoil the symmetries of the background we will consider only gauge transformations of the form $\xi_\mu = (\xi_t, \xi_r, \xi_\psi, 0, \frac{\cos\theta}{2}\xi_\psi)$ with $\xi_{t,r,\psi}$ being functions of $t$ and $r$ only.
By choosing
\begin{flalign}
\xi_{t} &= -\frac{f^2}{2r\Omega}\left[ h_{r\psi} +\left(1-\frac{2r^2(f^2-\Omega^2h^2)}{h^2f^2}\right)\frac{\xi_\psi}{r} 
+ \frac{\partial_r\xi_\psi}{2} \right]\,, \nonumber\\
\xi_{r} &= -\int \left[ 2h_{tr} + \frac{2}{r}\xi_t + \partial_r\xi_t -2\frac{(\ell^2+2r^2)}{r \ell^2 f^2}(\xi_t+\Omega\xi_\psi) \right] dr, \nonumber\\
\xi_{\psi} &= -(r^2+r^{-2}) \int_1^r \frac{J(t,\tilde{r})\,(\tilde{r}^2-\tilde{r}^{-2})}{8\tilde{r}} d\tilde{r} \nonumber\\
   & \qquad\qquad + (r^2-r^{-2}) \int_1^r \frac{J(t,\tilde{r})\,(\tilde{r}^2+\tilde{r}^{-2})}{8\tilde{r}} d\tilde{r}\,,
\label{gauge_transf}
\end{flalign}
with
\beq
J(t,r) &=& \frac{8r^3\Omega}{f^2}\left(h_{tr}+\Omega\, h_{r\psi} -\frac{2g^2}{r}\partial_t h_{\theta\theta}\right)  \nonumber\\
   && \qquad -\frac{8r^3-2r h^2}{h^2}h_{r\psi} -2r^2\partial_r h_{r\psi} \,,
\eeq
we manage to eliminate the components $h_{tr}, h_{r\psi}, h_{r\phi}$ and  $h_{\theta\theta}$, thus bringing the metric perturbation to the form~\eqref{eq:gaugefix}.

\bigskip
The gauge fixing we have just performed still leaves some {\em residual} gauge freedom: one can still make a coordinate transformation defined by 
$\xi_\mu^{\rm res} = (\zeta_t,0,\zeta_\psi,0,\frac{\cos\theta}{2}\zeta_\psi)$,
with
\beq
\zeta_t(t,r)      &=&  \omega_1(t)r^2 + \omega_2(t)r^{-2}\,, \\
\zeta_\psi(t,r) &=&  -a\left(1+\frac{r^2}{\ell^2}\right)\omega_1(t) + \frac{r^4+\ell^2r^2-2M\ell^2}{2 a M\ell^2 r^2}\omega_2(t)\,. \nonumber
\eeq
This leaves all the components unaltered except for $h_{tt}$ and $h_{t\psi}$ which transform as
\beq
h_{tt}     &\rightarrow&  h_{tt} -a\left(1+\frac{r^2}{\ell^2}\right)\omega_1'(t) + \frac{r^4+\ell^2r^2-2M\ell^2}{2 a M\ell^2 r^2}\omega_2'(t)\,, \nonumber\\
h_{t\psi} &\rightarrow&  h_{t\psi} +\frac{r^2}{2}\omega_1'(t) +\frac{1}{2r^2}\omega_2'(t)\,.
\label{residual_freedom}
\eeq
%

\bigskip
Several of the equations between~\eqref{gauge_transf} and~\eqref{residual_freedom} break down in the nonrotating limit, where $a=0$. This case must be dealt with separately, but it is naturally simpler. Assuming spherical symmetry of the background and of the perturbation, the only nonvanishing components of $h_{\mu\nu}$ are $\{h_{tt}, h_{tr}, h_{rr}, h_{\psi\psi}, h_{\psi\phi}, h_{\theta\theta}, h_{\phi\phi} \}$. However, the latter three components are determined by $h_{\psi\psi}$:
\be
h_{\psi\phi} = \frac{\cos\theta}{2}h_{\psi\psi}\,, \quad\;\;
h_{\theta\theta} = \frac{h_{\psi\psi}}{4}\,, \quad\;\;
h_{\phi\phi} = \frac{h_{\psi\psi}}{4}\,.
\ee

By making a gauge transformation of the form $\xi_\mu = (\xi_t, \xi_r, 0, 0, 0)$, with $\xi_{t,r}$ given by
\begin{flalign}
\xi_{t} &= \frac{r^4+\ell^2 r^2 - 2M \ell^2}{r^2} \int \left[ \frac{\ell^2r^3\, \partial_t h_{\psi\psi}}{\left( r^4+\ell^2 r^2 - 2M \ell^2 \right)^2} \right. \nonumber\\
   & \qquad\qquad - \left. \frac{2r\, h_{tr}}{r^4+\ell^2 r^2 - 2M \ell^2 } \right]dr\,, \nonumber\\
\xi_{r} &= - \frac{\ell^2 r \, h_{\psi\psi}}{r^4+\ell^2 r^2 - 2M \ell^2}\,,
\label{gauge_transf_Schw}
\end{flalign}
we can eliminate $h_{tr}$ and $h_{\psi\psi}$ and thus in the nonrotating case we can gauge fix the metric perturbation to take the form
\be
h_{\mu\nu}= \textrm{diag} \{ h_{tt}(t,r) , h_{rr}(t,r) , 0, 0, 0 \}\,.
\ee

There is still some residual gauge freedom left. Namely, by performing a coordinate transformation defined by $\xi_\mu^{\rm res} = \big(\frac{r^4+\ell^2r^2-2M\ell^2}{r^2}\omega_0(t),0,0,0,0\big)$, with $\omega_0(t)$ being a generic function of time, the $\{rr\}$ component remains unaltered while the $\{tt\}$ component changes as
\be
h_{tt}    \rightarrow  h_{tt}  + \frac{r^4+\ell^2r^2-2M\ell^2}{r^2}\omega_0'(t)\,.
\ee
This exhausts the amount of gauge freedom initially present in the metric perturbation tensor.

\begin{widetext}

\section{Equations governing linear perturbations in the rotating case
\label{sec:equations}}

The spherically symmetric case $a=L_\psi=0$ is a very special particular case in which one can solve the perturbation equations exactly. In this appendix we consider the more general case $a\neq0$, $L_\psi\neq0$. 

Plugging the metric perturbation ansatz into the linearized Einstein equations one finds that seven of these equations are automatically satisfied, leaving 8 nontrivial equations. Upon partial integration the full system reduces to just four independent equations. Specifically, we get a decoupled PDE for the component $h_{\psi\psi}$,
\begin{flalign}
&\frac{3\ell^2 (r^4+2 a^2 M)^2}{ r^6 + \ell^2r^4 + 2M(a^2 - \ell^2)r^2 + 2 a^2 M \ell^2 } \partial_t^2 h_{\psi\psi} \;
- \; \frac{3(r^4+2a^2 M) \left(r^6 + \ell^2 r^4 + 2M (a^2-\ell^2) r^2 + 2 a^2 M \ell^2 \right)}{\ell^2 r^4} \partial_r^2 h_{\psi\psi}  \nonumber\\
-& \frac{3(r^4+2 a^2 M) \left( 3 r^{10} - 3 \ell^2 r^8 + 18 \ell^2 M r^6 - 16 a^2 \ell^2 M r^4 + 20a^2M^2(a^2 -\ell^2) r^2 + 12 a^4 \ell^2 M^2 \right)}{\ell^2 r^5 (3 r^4+2 a^2 M)} \partial_r h_{\psi \psi}  \nonumber\\
+& \frac{12 (r^4+2 a^2 M) \left( 3 r^8 + 6 \ell^2 r^6 + 2M(2 a^2+3 \ell^2) r^4 + 20 a^2 \ell^2 M r^2 - 4 a^2 M^2 (a^2-\ell^2) \right)}{\ell^2 r^4 (3 r^4+2 a^2 M)} h_{\psi \psi} \nonumber\\
+& \frac{3072\pi\,m_0 a M (r^4+2 a^2 M) (2 a E-3 L_{\psi})}{r^2(3 r^4+2 a^2 M)} \Theta(r-R) \;
+ \; \frac{64 \pi\, m_0(r^4+2 a^2 M)}{\ell^2 r^7} \chi_1(t,r) \delta(r-R) = 0\,,\label{hpsipsi}
\end{flalign}
a coupled ODE for $h_{t\psi}$ and $h_{\psi\psi}\,$, which can be suggestively expressed as
\beq
\partial_r \left( \frac{1}{r^3} \partial_r \left(r^2 h_{t\psi}\right)\right) &=& \frac{1}{r^3(r^4+2a^2M) (3 r^4+2 a^2 M) (r^6+\ell^2r^4+2M(a^2-\ell^2)r^2+2a^2M \ell^2)} \nonumber\\
&\times& \Big\{ 2a M r^2 (r^4-2 a^2 M) (r^6+\ell^2r^4+2M(a^2-\ell^2)r^2+2a^2M \ell^2)\, \partial_r^2 h_{\psi \psi} \nonumber\\
&&\;\; - \frac{2a M r}{ (3 r^4+2 a^2 M) } \Big[-3 r^{14}+9 \ell^2 r^{12}+14M(a^2-3 \ell^2) r^{10}+46 a^2 M\ell^2 r^8 \nonumber\\
&&\qquad\qquad -4 a^2 M^2(5 a^2-6 \ell^2) r^6 - 4 a^4 M^2 \ell^2 r^4+8 a^4 M^3 (a^2-\ell^2) r^2+8 a^6 M^3 \ell^2 \Big] \partial_r h_{\psi \psi}  \nonumber\\
&&\;\; - \frac{8a M}{ (3 r^4+2 a^2 M) }  \Big[3 r^{14}+9 \ell^2 r^{12}-6M (a^2-3 \ell^2) r^{10}+26 a^2 M \ell^2 r^8 \nonumber\\
&&\qquad\qquad -4 a^2 M^2 (7 a^2-6 \ell^2) r^6-20 a^4 M^2 \ell^2 r^4-8 a^4 M^3 (a^2-\ell^2) r^2-8 a^6 M^3 \ell^2 \Big] h_{\psi \psi} \nonumber\\
&&\;\; - \frac{4096 \pi \, m_0a^2 M^2 \ell^2 (2 a E - 3L_{\psi})\, r^8}{(3 r^4+2 a^2 M)} \Theta(r-R) 
+  \frac{64\pi \, m_0 r^3}{3} \chi_2(t,r) \delta(r-R) \Big\}
\,,\label{htpsi}
\eeq
and two constraint equations determining $h_{rr}$ and $h_{tt}\,$,
\beq
h_{rr} &=& \frac{ r^6 + \ell^2r^4 + 2M(a^2 - \ell^2)r^2 + 2 a^2 M \ell^2}{\ell^2 r (3 r^4+2 a^2 M)} \partial_r h_{\psi\psi} 
- \frac{2r^6 + \ell^2r^4 - 2 a^2 M \ell^2}{\ell^2 r^2 (3 r^4+2 a^2 M)} h_{\psi\psi} \nonumber\\
&& + \frac{64 \pi\, m_0 \left(E r^4 +2 a M(a E - L_{\psi})\right)}{r^2(3 r^4+2 a^2 M)} \Theta(r-R)\,, \label{hrr}\\
&&\; \nonumber\\
h_{tt} &=& - \frac{r^6+\ell^2r^4+2M(a^2-\ell^2)r^2+2a^2M \ell^2}{2 \ell^2 r (3r^4+2 a^2 M)} \partial_r h_{\psi\psi} 
- \frac{2r^6+\ell^2r^4+2 a^2 M (r^2+\ell^2)}{\ell^2 r^2 (3r^4+2 a^2 M)} h_{\psi\psi}\nonumber\\
&&- \frac{r^6+\ell^2r^4+2M(a^2-\ell^2)r^2+2a^2M \ell^2}{4 a M \ell^2r} \partial_r h_{t\psi} 
+ \frac{r^6+\ell^2r^4-2M(a^2+\ell^2)r^2-2a^2M \ell^2}{2 a M \ell^2 r^2} h_{t\psi} \nonumber\\
&&+ \frac{16 \pi  m_0 \left(4 a M \ell^2 E r^2 + L_{\psi} \left[3 r^6 + 3 \ell^2 r^4 + 2M (a^2-3 \ell^2) r^2 + 2 a^2 M \ell^2\right]\right)}{a M \ell^2 (3r^4+2 a^2 M)} \Theta(r-R)\,, \label{htt}
\eeq
where $R=R(t)$ is to be understood as a function of $t$, and for convenience we have introduced in Eqs.~\eqref{hpsipsi} and~\eqref{htpsi} the following functions multiplying the delta distributions:
\beq
\chi_1(t,r) &\equiv&  -4 a^2 M \ell^2 E r^2 + L_{\psi} \left[6aM\ell^2r^2 - r^2 (3 r^4+2 a^2 M)\frac{ (R^4+\ell^2 R^2-2M \ell^2) L_{\psi}+2 a M \ell^2 E}{E R^4+2 a M (a E-L_{\psi})}\right] \nonumber\\
&& + \frac{4a^2 M \ell^6 r^6 \left(E R^4+2 a M (a E-L_{\psi})\right) R'(t)^2}{\left(r^6+\ell^2 r^4+2M (a^2-\ell^2) r^2+2 a^2 M\ell^2\right) \left(R^6+\ell^2 R^4+2M(a^2-\ell^2)R^2+2 a^2 M \ell^2\right)}\,, \\
\; \nonumber\\
\chi_2(t,r) &\equiv& 4 a M \ell^2 E r^2 (3 r^4+10 a^2 M) +\frac{4 a M \ell^4 r^6 (3 r^4+2 a^2 M)}{r^6+\ell^2r^4+2M(a^2-\ell^2) r^2+2 a^2 M \ell^2}R'(t)^2 \dot{T}(\tau) \nonumber \\
&& + L_{\psi} \Big[3 \Big(3 r^{10}+3 \ell^2 r^8+(8 a^2 M-6 \ell^2 M) r^6+8 a^2 M \ell^2 r^4+4 a^2 M^2 (a^2-5 \ell^2) r^2+4 a^4 M^2 \ell^2\Big) \nonumber \\
&&\qquad\;\; + 4 a M \ell^2 r^2 (3 r^4+2 a^2 M) \Psi'(t) \Big] \,.
\eeq
In writing these equations we have dropped a couple of integration constants. This is justified since we are looking for perturbations that are sourced by the test shell.
Thus, the procedure is to (i) find a solution of the PDE for the component $h_{\psi\psi}$, (ii) replace it in the coupled ODE and determine the solution for $h_{t\psi}$, and (iii) obtain components $h_{rr}$ and $h_{tt}$ from the constraints~\eqref{hrr} and~\eqref{htt}.

\bigskip

\end{widetext}

\section{Strategy to solve PDEs with distributional sources
\label{sec:checks}}

The equation~\eqref{hpsipsi} is a second order nonhomogeneous PDE, with source terms proportional to distributions, namely the Heaviside function and the Dirac $\delta$-function. It is convenient to recast the problem as a homogeneous PDE with prescribed boundary conditions dictated by the nonhomogeneous terms. This can be done as follows.

Plugging in solutions of the form~\eqref{eq:ansatz} the differential equation will take the general form
\beq
S(R(t),r)\,\Theta(r-R(t)) &+& P(R(t),r)\,\delta(r-R(t)) \nonumber\\
&& \!\!\!\!\!\!\!\!\!\!\!\!\!\!\!\!\!\!\!\!\!\!\! + Q(R(t),r)\,\delta'(r-R(t)) = 0\,.
\eeq
One concludes immediately that the coefficient of the Heaviside function must vanish since it is the only term with support away from $r=R(t)$. By integrating in $r$ or in $R$ we obtain three other constraints that may be regarded as boundary conditions. The final result is
\beq
S(R,r) &=& 0\,, \\
Q(R,R) &=& 0\,, \label{BC1}\\
P(R,R) - \partial_r Q(R,R) &=& 0\,, \label{BC2}\\
P(R,R) + \partial_R Q(R,R) &=& 0\,. \label{BC3}
\eeq
In fact, assuming $Q$ to be an analytic function in the variable $r$, Eq.~\eqref{BC3} can be derived from Eqs.~\eqref{BC1} and~\eqref{BC2}, so it need not be imposed separately.



\begin{thebibliography}{99}
\bibitem{footnote1}
To be precise, the number of independent angular momenta is equal to the rank of the rotation group, which is in correspondence with the number of orthogonal rotation planes.
\bibitem{footnote2}
For equal angular momenta this symmetry is enhanced from $U(1)^2$ to $U(2)$.
\bibitem{footnote3}
The restriction to equally spinning backgrounds was instrumental in rendering the linearized perturbation equations amenable to analytic treatment. Relaxing this assumption, one generically obtains additional independent metric perturbation components, a larger set of equations to solve and explicit dependence on polar angles. Nevertheless, the general approach we have followed is still applicable in these cases.
\end{thebibliography}
\end{document}